
\normalbaselineskip = 12 pt
\magnification = 1200
\hsize = 15 truecm \vsize = 22 truecm \hoffset = 1.0 truecm
\medskip
\centerline{{\bf DESCRETE SPECTRUM OF DELOCALIZED STATES}}
\medskip
\centerline{{\bf IN ONE-DIMENSIONAL RANDOM POTENTIAL}}
\medskip
\centerline{{\bf M. Yu. Lashkevich}}
\medskip
\centerline{Landau Institute for Theoretical Physics, Academy of Sciences,}
\centerline{Kosygina 2, 117334 Moscow, Russian Federation}
\medskip
\par
It is well known$^{1,2}$ that in one-dimensional disordered system
all states of electrons (or any other exitations) are localized.
In this letter it is shown that delocalized states exist in a rather
broad class of of simple models, but a set of delocalized states is
not too great, and it does not contradict this theorem in more precize
form$^{2,3}$. These systems can be considered as filters in energy
for electrons.
\par
Discuss an accidental infinite chain of fragments of two types $F_1$ and
$F_2$ with lengths $a_1$ and $a_2$ respectively. Through the chain an electron
spreads. The potential of fragment $F_i$ is $V_i(x)$, where $x$ is a
coordinate with respect to the left end of the fragment, $0\leq x\leq a_i$.
Change slightly the system: embed between any two neibouring fragments
a plot $F_0$ of constant potential $V_0$ with length $b$. If $b$ tends to
zero, the stationary states of the electron tend evidently to those of
initial system independent on $V_0$.
\par
Consider a fragment $F_i$ embedded in media with constant potential $V_0$.
If a wave falls on the fragment from the left, the wave function has the
form
$$
\psi(x)=e^{ikx}+\alpha_i(E,V_0)e^{-ikx},{\hbox{ if }}x<0;
$$
$$
\psi(x)=\beta_i(E,V_0)e^{ikx},{\hbox{ if }}x>a,\eqno(1)
$$
where $E$ is energy of electron, $k={1\over\hbar}\sqrt{2m(E-V_0)}$;
$\alpha_i(E,V_0)$ and $\beta_i(E,V_0)$ are amplitudes of reflected and
passed waves respectively. Solution of two equations with two
variables
$$
\alpha_1(E,V_0)=0,
$$
$$
\alpha_2(E,V_0)=0\eqno(2)
$$
must consist of separate points $(E^{(n)},V_0^{(n)})$, $n=1,2,...$
on the plane $(E,V_0)$. But eqs.(2) mean that an electron with energy
$E^{(n)}$ in the media with potential $V_0^{(n)}$ does not reflect from
any fragment $F_1$ or $F_2$, which are embedded in any region of media
in any quantity and order.
\par
Come back to the system of fragments $F_1$, $F_2$ and $F_0$. Suppose
that $V_0=V_0^{(n)}$ and energy of electron is $E^{(n)}$. If in one
of the fragments $F_0$ the wave function $\psi_n(x)$ is equal to
$\exp{(ikx)}$, then on any other fragment $F_0$ the wave function is
equal to $\exp{(ikx+\varphi)}$, where $\varphi$ is a real number
(depending on fragment), because $|\beta_1(E^{(n)},V_0^{(n)})|^2=
|\beta_2(E^{(n)},V_0^{(n)})|^2=1$. The wave function $\psi_n(x)$ is
delocalized. If $b\rightarrow 0$, the state with energy $E^{(n)}$ remains
delocalised, but dependance on auxilar quantity $V_0$ disappears.
States with all energies $E^{(n)}$ become delocalized. We see that
these descrete delocalized states must exist not for some special
potentials $V_1(x)$ and $V_2(x)$, but for a broad class of them.
\par
Consider an example. In the case $V_i(x)=V_i=const$ there are two
sets of delocalized levels
$$
E^{(n,1)}={{\hbar^2\pi^2n^2}\over{2ma_1^2}}+V_1,\quad
E^{(n,2)}={{\hbar^2\pi^2n^2}\over{2ma_2^2}}+V_2,\quad
n=1,2,\cdots.\eqno(4)
$$
It is interesting to notice that wave length of $\psi_{n,i}(x)$ on the
fragment $F_i$ is equal to $a_i/2n$.
\par
It seems to be probable that delocalized
states are not stable with respect to
perturbation of model, and they become localized states of large size
under the influence of week perturbation, if perturbed system cannot
be consider as a chain of two types of fragments.
\par
At last, note that the condition (2) can be rewritten in the form
which is independent from $V_0$:
$$
\alpha_1(E)=\alpha_2(E).\eqno(5)
$$
One can say that for delocalized states the chain looks as a periodical
up to an accidental phase $-i\ln{\beta_i}$, which is not essential,
because there are no roundabout ways. In other words eigenvectors of
both transition matrices coincide. Hence, there are no contradictions
with the theorem about localization$^{2,3}$.
\par
The author is greatful to V. T. Bublik for support and A. V. Ruban for
discussions.
\medskip
{\bf References}
\par
1. N.F.Mott and W.D.Twose, $Adv.$ $Phys.$ {\bf 10} (1961) 107
\par
2. J.M.Ziman, Models of Disorder, Cambridge University Press,
Cambridge, 1979
\par
3. H.Matsuda and K.Ishii, $Prog.$ $Theor.$ $Phys.$, $Supp.$,
{\bf 45} (1970) 56
\end